\documentstyle{article}

\def\ove#1{\over \,\, #1 \,\, }
\def\frace#1#2{\frac{#1}{\,\, #2 \,\, }}
\newcommand{\be}{\begin{equation}}
\newcommand{\ee}{\end{equation}}
\newcommand{\ba}{\begin{eqnarray}}
\newcommand{\ea}{\end{eqnarray}}

\newcommand{\LA}{\Lambda | \Sigma}
\newcommand{\ALA}{{\overline \Lambda} | {\overline \Sigma}}
\newcommand{\XI}{\Xi}
\newcommand{\AXI}{\overline \Xi}
\newcommand{\OM}{\Omega}
\newcommand{\AOM}{\overline \Omega}

\newcommand{\NAQ}{{\overline q}}
\newcommand{\AQ}{{\overline q}}
\newcommand{\aq}{{\overline Q}}
\newcommand{\NQ}{q}
\newcommand{\NS}{s}
\newcommand{\NAS}{{\overline s}}
\newcommand{\AS}{{\overline s}}
\newcommand{\as}{{\overline S}}
\newcommand{\PR}{N}
\newcommand{\APR}{\overline N}
 
\newcommand{\CP}{C_p}
\newcommand{\CL}{C_{\Lambda}}
\newcommand{\CX}{C_{\Xi}}
\newcommand{\CO}{C_{\Omega}}

\newcommand{\CAP}{C_{\overline p}}
\newcommand{\CAL}{C_{\overline \Lambda}}
\newcommand{\CAX}{C_{\overline \Xi}}
\newcommand{\CAO}{C_{\overline \Omega}}
 
\newcommand{\CPI}{C_{\pi}}
\newcommand{\CK}{C_{K}}
\newcommand{\CAK}{C_{\overline K}}
 
\newcommand{\AAO}{a_{\overline \Omega}}
\newcommand{\AO}{a_{\Omega}}
\newcommand{\AAX}{a_{\overline \Xi}}
\newcommand{\AX}{a_{\Xi}}
\newcommand{\AAL}{a_{\overline \Lambda}}
\newcommand{\AL}{a_{\Lambda}}
\newcommand{\AAP}{a_{\overline p}}
\newcommand{\AP}{a_{p}}
 
\newcommand{\NP}{{p}}
\newcommand{\NAP}{{\overline p}}
\newcommand{\NLA}{{\Lambda} | {\Sigma}}
\newcommand{\NAL}{{\overline \Lambda} | {\overline \Sigma}}
\newcommand{\NX}{{\Xi}}
\newcommand{\NAX}{{\overline \Xi}}
\newcommand{\NO}{{\Omega}}
\newcommand{\NAO}{{\overline \Omega}}
 
\newcommand{\NPI}{{\pi}}
\newcommand{\NK}{{K}}
\newcommand{\NAK}{{\overline K}}
 
\newcommand{\BM}[2]{ b_{#1} \, b_{#2} \, {#1} \, {#2} }

\begin{document}

\normalsize
\parindent0mm
 
 
\title{\bf Quark liberation and coalescence at CERN SPS}
 
\author{{\it J. Zim\'anyi, T.S. Bir\'o, T. Cs\"org\H o and P. L\'evai} \\ 
	MTA KFKI RMKI, \\
	H - 1525 Budapest 114, P.O.Box 49, Hungary }
\date{August 2, 1999 }

\maketitle


\abstract{The mischievous linear coalescence approach
to hadronization of quark matter is shown to 
violate strangeness conservation in strong interactions.
The simplest  correct quark counting 
is shown to coincide with the  non-linear
{\it al}gebraic {\it co}alescence {\it r}ehadronization  model, ALCOR. 
The non-linearity of the ALCOR model is shown to
cancel from its simple predictions for the
relative yields of (multi-)strange baryons.
We prove, model independently, that quark degrees of freedom 
are liberated  before hadron formation 
in 158 AGeV  central Pb + Pb collisions at CERN SPS. 
}
 
{PACS number(s): 12.38.Mh, 13.87.Fh, 24.85.+p}

 
\section{Introduction}
It is a deep rooted desire to explain complicated experimental
observations with simple and transparent models, understandable
in laymen's terms. 
This very acceptable ambition inspired a recent publication ~\cite{Bial},
where an attempt was made to explain the
relations between the multiplicities of different  strange baryons
produced in heavy ion reactions with the help of linear
coalescence rehadronization model. This  
model was based on
simple quark counting and elementary probability estimates. 
However, the considerations in Ref.~\cite{Bial}
lead to 
a clear violation of strangeness conservation in strong interactions:
In order to explain the data in laymen's terms, Bialas had to assume 
in heavy ion reactions the number produced of strange quarks, $  \NS $ is not 
equal with the number of produced anti-strange quarks, $ \NAS $. 

We show here that this unacceptable assumption was related to the neglected
requirement that all quarks have to hadronize, as no free quarks are observed
in nature. This necessity leads to a non-linearity
even in the simplest algebraic rehadronization model (see Ref.~\cite{bz}). 
Properly taking into account the non-linear competition
for quarks by the various hadron forming coalescence channels
is a natural way to  correct the  linear coalescence
treatment in the simplest possible manner.
This leads to
 the normalized, non-linear {\it al}gebraic {\it co}alescence 
{\it r}ehadronization model,
ALCOR \cite{ALCOR},
which automatically takes care of the conservation of strangeness in strong 
interactions as well.
 
\section{Linear coalescence in rehadronization?}
In the linear coalescence rehadronization model Bialas assumed, 
that the number of produced particles is proportional 
to the product of the numbers of constituent 
particles within the reaction volume. 
In the following we denote
the {\it number} of particles by the particle symbols, 
in particular, $\NQ$ stands for the number of light (up and down)
quarks.
Linear coalescence yields the following relations:
\ba
    \NP  &=& \AP\, \NQ^3, \label{e:lic1} \\
    \NLA &=& \AL\, \NQ^2 \, \NS, \\ 
    \NX  &=& \AX\, \NQ^{\phantom{2}} \, \NS^2 ,\\
    \NO  &=& \AO\, \NS^3,
\ea
where $\NLA = \Lambda + \Sigma$ stands for
the total number of strange baryons  that contain a single strange
quark. 

\newpage
Similarly, for the anti-baryons one obtains 
\ba
    \NAP  &=& \AAP\,\, \NAQ^{\, 3}, \\
    \NAL  &=& \AAL\,\, \NAQ^{\, 2}\,\, \NAS, \\
    \NAX  &=& \AAX\,\, \NAQ^{\phantom{\, 2}}\, \NAS^{\, 2}, \\ 	
    \NAO  &=& \AAO\,\, \NAS^{\, 3}.  \label{e:licu}
\ea
Furthermore, Ref.~\cite{Bial} also assumed, that 
the coefficients of proportionality for particles and their antiparticles
are equal:  $ \AAO = \AO $,  etc. As the coefficients $\AAO, \AO$ describe
effectively the (particle anti-particle symmetric) likelihood 
that a multi-quark bound state is formed once the constituents are given,
this seems to be a very reasonable model at first sight.

This linear model is attractive not only because of its simple formulation
but also because of its simple predictions, namely that the unknown
$a$ coefficients cancel from the anti-baryon to baryon
ratios. This leads to the following equations:
\ba
    {\NAP \ove \NP }  &=& \left[{\NAQ \ove \NQ }\right]^3, \label{e:ao1} \\
    {\NAL \ove \NLA }  &=& \left[{\NAQ \ove \NQ }\right]^2 \,
				\left[{\NAS \ove \NS }\right], \\
    {\NAX \ove \NX }  &=& \left[{\NAQ \ove \NQ }\right]^{\phantom{2}} \,
			    \left[{\NAS \ove \NS }\right]^2, \\ 	
    {\NAO \ove \NO }  &=& \left[{\NAS \ove \NS }\right]^3. \label{e:aow}
\ea

In order to explain the value of 

\be
{\AOM \ove \OM} =  0.383 \pm 0.081, 
\ee 
as measured in Pb +Pb reactions at CERN SPS energies,
Bialas had to assume \ \  $ \NS \neq \NAS $ as a consequence of
eq.~(\ref{e:aow}) of his linear model,
{\it which is a  clear violation  of strangeness conservation 
in strong interactions.}

What is the origin of this contradiction?

The linear coalescence model, as defined  in eqs. (3) and (4)
is a good approximation {\it only if 
the composite particles use up  a small fraction} of the
constituents.
This is the case e.g. in the coalescence treatment of
the deuteron $(pn)$ or triton $(pnn)$ formation from the gas of protons
and neutrons.
In that situation
most particles will remain in nucleon ($p$ or $n$) state. Thus
the competition is negligible among the tritons and the deuterons 
for the building block nucleons in that case.
 
In the case of hadronization of a quark matter {\it all} of the constituent
particles (the quarks) have to be placed into composite particles,
namely into colorless hadrons. No free quarks are observed in Nature.
{\it This is the very essence of color confinement.}
Thus, during the hadronization process, 
a fixed number of strange quarks
must be distributed among hyperons and anti-K mesons.
The anti-strange quarks have to be distributed
among anti-hyperons and K mesons. [ Note that K meson is
a $(q {\overline s})$ bound state, while
the anti-K meson is  a $ (\overline q s )$ bound state.]
However, in heavy ion collisions we have incoming nuclei in the initial
state, that contain nucleons, formed by constituent quarks, only. 
As quarks are produced in quark  antiquark pairs,
the number of quarks in the final state is larger than the
number of anti-quarks. Hence  we must have more K mesons
than anti-K mesons.

 {The processes creating different hadrons
 are not independent, they compete with each other,
 {\it contrary} to the basic assumption in the linear coalescence
model.}
 
The redistribution can be counted for by introducing normalization
factors, $b_s, b_{\overline s}, b_q, b_{\overline q}$.
These $b_i$ normalization factors are not free parameters!
They are determined from the requirement that all quarks must be 
recombined into hadrons during the hadronization process,
as described first in the { ALCOR} model, Ref.~\cite{ALCOR}.

\section{The ALCOR approach}
The non-linear coalescence equations for the formation of
 quark-antiquark clusters during rehadronization read as:
\ba
    \NP  &=& \CP \, b_q^3 \, q^3, \label{coal1}  \\ 
    \NLA &=& \CL \, b_q^2 \, b_s \, q^2 \, s,    \\
    \NX  &=& \CX \,  b_q^{\phantom{2}} \, b_s^2 \, q^{\phantom{2}} \, s^2, \\
    \NO  &=& \CO \, b_s^3 \, s^3, \\
    \NAP  &=& \CAP \,  b_\AQ^3 \, \AQ^3, \\ 
    \NAL  &=& \CAL \, b_\AQ^2 \, b_\AS \, \AQ^2 \, \AS,    \\
    \NAX  &=& \CAX \, b_\AQ^{\phantom{2}} \, b_\AS^2 \, 
			\AQ^{\phantom{2}} \, \AS^2,
			\label{coal2}    \\
    \NAO  &=& \CAO \, b_\AS^3 \, \AS^3.
\ea
The meson yields are determined similarly, 
\ba
    \NPI^d  &=& \CPI \, \BM{q}{\overline q},  \\ 
    \NK  &=& \CK \, \BM{q}{\overline s},     \\ 
    \NAK  &=& \CAK \, \BM{\overline q}{s},  \\
    {\eta} &=& C_{\eta} \, \BM{s}{\overline s}.  \label{coal3}
\ea
The normalization coefficients $b_q$, $b_s$, $b_{\overline q}$ and 
$b_{\overline s}$ are determined uniquely
by the requirement, that {\it the number of the constituent
quarks do not change during the hadronization} ---
which is the basic assumption for all quark counting methods:
 
\ba
 s &=& 3 \, \NO + 2 \, \NX +  \NLA +
       {\overline K} +  {\eta}, \label{e:cons1}\\ 
 {\overline s}  &=& 3 \, \NAO + 2 \, \NAX +  \NAL +
       {K} +  {\eta}, \\ 
 q &=& 3 \, \NP +  
	2 \, \NLA + 
	 \NX +  {K} +  \NPI^d,  \\
 {\overline q}  &=& 
	3 \, \NAP +  
	2 \, \NAL +
	\NAX + 
      {\overline K}
    + \NPI^d  \ . \label{e:cons4} 
\ea
Here $\NPI^d $ is the number of directly produced $(q{\overline q})$ states.
( Note that other particle
symbols also stand for the number of directly produced particles
of a given type, and care must be taken when comparing the ALCOR predictions
with data, especially regarding the corrections of particle numbers
for the feed-down of resonance decay contributions.)

Similar, non-linear coalescence equations with constraints
of using up all the coalescing particles were proposed 
in a simpler form first in Ref.~\cite{bz}
in the so called combinatoric break-up model, 
assuming a constant coalescence coefficient for all the baryons and another
one for all the mesons. Later the ALCOR model was formulated as presented
above in a simple  form.  ALCOR equations were 
solved numerically in Refs.~\cite{ALCOR,ALC97,ALC99}, using,
as input values to the coalescence equations,
the number of quarks as predicted by Monte Carlo event generators.
The ALCOR equations were derived from a set of rate equations
in Ref.~\cite{ALC96}, in the 
sudden approximation.

Substituting eqs.(\ref{coal1} - \ref{coal3}) into the constraints 
given by eqs.~(\ref{e:cons1}-\ref{e:cons4}) one
obtains  a set of non-linear equations
for the $b_i$ normalization constants.
{\it However, one can predict some relations even without solving these
set of nonlinear equations}.
It turns out that 
the particle-antiparticle ratios can be expressed in terms of
the effective number of quarks even in the non-linear ALCOR model,
similarly as it was done by  Bialas in the linear coalescence model.

Let us denote by upper case $Q$, $S$ the reduced, effective
number of up+down  and strange quarks, in contrast to the lower case
$q$ and $s$, that stand for the total number of (up+down) and
strange quarks before the hadronization:
\ba
    Q &=& b_q \, q,\\ 
    {\overline Q} &=& b_{\overline q} \, {\overline q },\\ 
    S &=& b_s \, s, \\ 
    {\overline S} &=& b_{\overline s} \, {\overline s }
\ea
Furthermore, one could make the usual assumption
that the $C$ coalescence  coefficients for  baryons 
are equal to that of the corresponding antibaryons,
\ba
   \CP &=& \CAP, \\ 
   \CL &=& \CAL, \\ 
   \CX &=& \CAX, \\ 
   \CO &=& \CAO.
\ea
With these notations, the following relations
are obtained in the non-linear ALCOR model: 
\ba
    {\NAP \ove \NP }  
			&=& 
			\left[{\aq \ove Q }\right]^3, \\
    {\NAL \ove \NLA }  &=& 
			\left[{\aq \ove Q }\right]^2 \,
				\left[{\as \ove S }\right], \\
    {\NAX \ove \NX }  &=& 
			\left[{\aq \ove Q }\right]^{\phantom{2}} \,
			    \left[{\as \ove S }\right]^2, \\ 	
    {\NAO \ove \NO }  &=& 
			\left[{\as \ove S }\right]^3,  \\
    {\NAK \ove \NK }  &=& 
			 {\aq\, S \ove Q \, \as }. \label{e:aoa}
\ea
These equations are formally similar to the equations of the linear treatment
displayed in eqs.~(\ref{e:ao1}-\ref{e:aow}), however, they are given
in terms of the effective number of quarks, $Q = b_q q$, that
are complicated, non-linear functions of the number of quarks $q, s$ 
available before the hadronization.
Thus one easily gets the following interesting relations:
\ba
    {\NAL \ove \NLA } &=& {\NAP \ove \NP }  \, 
    			\left[ {\NK \ove \NAK } \right], \label{e:i1}\\
    {\NAX \ove \NX }  &=& {\NAP \ove \NP }  \, 
    			\left[ {\NK \ove \NAK } \right]^2, \label{e:i2}\\
    {\NAO \ove \NO }  &=& {\NAP \ove \NP }  \, 
    			\left[ {\NK \ove \NAK } \right]^3.\label{e:i3}
\ea
These are the relations among the  ratios of the observable
number of particles that should be satisfied if 
the particle production in some reaction proceeds via
algebraic recombination of the independent valence quarks.
This way a direct relation is obtained in a self-consistent manner,
that connects the presence/absence of independent, initial quarks 
with the observable yields of hadrons. This relation is to a large
extent model-independent (i.e. independent of the initial quark
content, independent from the values of the coalescence coefficients
$C$ and independent from the values of the non-linear
renormalization factors $b_i$).   Only two physical assumptions
enter these model-independent eqs.~(\ref{e:i1}-\ref{e:i3}):
that the rehadronization process is sudden and that 
the valence quarks are
available in unbound states before  the hadronization.

One may evaluate the following measurable  numbers:
\ba
	d_{\Lambda} & = & 
    		{\NAL \ove \NLA } \, {\NP \ove \NAP }, \\
	d_{\Xi} & = & 
    		\left[{\NAX \ove \NX } \, {\NP \ove \NAP }\right]^{1/2}, \\
	d_{\Omega} & = & 
    		\left[{\NAO \ove \NO } \, {\NP \ove \NAP }\right]^{1/3},\\ 
      	d & = & 
    			\left[ {\NK \ove \NAK } \right]. 
\ea
 If hadronization proceeds via a sudden recombination of quarks.
then all these $d$ numbers should be equal, 
$d = d_{\Lambda}  = d_{\Xi}  =  d_{\Omega}$.
 
The experimental values for the anti-baryon to baryon number ratios
are determined in Refs.~\cite{EXP1,EXP2,EXP3} as follows:
\ba
 \frac {\APR}{\,\, \PR\,\, } 
	&=&  0.070 \pm 0.010 ,\\ 
 \frac {\ALA} {\,\, \LA \,\,} &=&  0.133 \pm 0.007 ,\\ 
 \frac {\AXI} {\,\, \XI \,\,} &=&  0.249 \pm 0.019 ,\\ 
 \frac {\AOM} {\,\, \OM \,\,} &=&  0.383 \pm 0.081 ,\\
 \frac {K^+} {\,\, K^-\,\, }  &=&  \,\, d \,\, = \,\, 1.80  \pm 0.2
\ea
which yields the following values for the $d$ factors of strange
baryons :
\ba
     d_{\Lambda} &=& 1.9  \pm 0.3,\\
     d_{\Xi}     &=& 1.89 \pm 0.15,\\
     d_{\Omega}  &=& 1.76 \pm 0.15.
\ea
These numbers show a very good agreement with $d$, 
a statistically acceptable, good agreement between the ALCOR model
and the experimental data in case of Pb + Pb reaction at CERN SPS.
This indicates, regardless of the details of the confinement
mechanism, that hadron production in
Pb +Pb reaction at CERN SPS  proceeds via a sudden and complete
coalescence of constituent quarks to hadrons.

Taking into account the conservation of strangeness in strong interactions,
which demands ${\overline s} = s, $ 
from the $ { \overline \Omega } / \Omega $ ratio we arrive at
\be
    \frac{ {\overline S}} { \,\,  S \,\, } =
    \frac{ b_{\overline s} } { b_s } =
    0.75 \pm 0.06.
\ee

Thus  we obtained agreement with the experimental
data without assuming $ s \neq {\overline s}.$
The deviation of $ \overline \Omega / \Omega $
from unity is caused by the difference in the
normalization factors $ b_s$ and $ b_{\overline s} $,
which can easily be understood:
there are more quarks then antiquarks in the initial
system, and thus more $ \overline s $ quarks are
used up in the $ K^+$ 
production
then $ s $ quarks in the $ K^- $ meson creation.
Thus less $\overline s $ remains for the $ \overline \Omega $
production then $ s $ quarks for the $ \Omega $
production.

In the time-dependent solution of confining rate equations
for the hadronization~\cite{ALCOR2}, conservation laws are ensured
by the structure of the coupled system of differential equations. 
The ALCOR model, as presented above, can be reobtained from a set of non-linear 
rate equations in the sudden approximation~\cite{ALC96}, in
the limit when the time of hadronization is very short.
Indeed, a detailed analysis of single particle spectra and two-particle
correlation data indicates a short duration, $\Delta \tau \approx 1.5$ fm/c,
for the production of final state hadrons at CERN SPS Pb + Pb reactions
~\cite{scl99}. This result excludes a long-lived, evaporative mixture of
quark-gluon plasma and hadronic phase, that would produce pions over 
an order of magnitude larger period. Thus the experimental data prefer
a sudden production of hadrons, a process that may happen to be 
 out of local thermal and chemical equilibrium.

Finally, let us point out, that the strangeness conservation
leads to the following relation:
\ba
s & =&   {\overline s},\\
\null\hspace{-0.4cm}    
 3\, \frac{\NO}{\,\, \NX \,\, } \, \frac{\NX}{\,\,\NLA\,\,} 
+ 2\,\frac{\NX}{\,\,\NLA\,\,}
+ \frace{{\overline K} }{\NLA} 
 + 1 
\!\! & = &  \!\!
 \frace{\NAL}{\NLA}\, \left[3\,\frace{\NAO}{\NAX}\,\frace{\NAX}{\NAL}
  +2 \, \frace{\NAX}{\NAL} 
+ \frace{K}{\NAL}
+ 1 
\right]. \label{totals} 
\ea
With the SPS Pb+Pb data \cite{EXP1,EXP2,EXP3} 
the left hand side of this equation is $2.57$,
while the right hand side is $2.66$. Thus the strangeness
conservation equation is also well fulfilled for this case.
The validity of eq.~(\ref{totals}) is an absolute measure of
strangeness conservation in strong interactions. 
If the experimental values do not satisfy eq.~(\ref{totals}),
then the detectors miss important parts of the particle
momentum distributions, or they have a serious 
systematic error. 

Further observations of Ref. \cite{Bial}, namely 
$  d_{\Lambda} =  d_{\Xi}   $  for the SPS
$^{32}S+ ^{23}S$ reactions, but $ d_{\Lambda} \neq  d_{\Xi}   $ 
for the SPS p+Pb reaction remain unchanged
with our analysis.
 
\section{ Conclusions} 
We clarified the
reason  why the linear coalescence 
model should not be applied to the hadroni\-za\-tion of 
 the quark  matter: it leads to a violation  
of strangeness conservation. 

We have shown, however, that this shortcoming can be corrected
if a  a nonlinear coalescence model is introduced, 
that takes into account the conservation of strangeness in
strong interactions and the fact that no free quarks are observed
in Nature. In particular, we have shown that
the solution of the ALCOR  hadronization model 
yields simple, parameter independent relations
in a consistent manner. 
These relations were surprisingly simple and similar to those obtained from
the linear coalescence picture. Data on strange particle production  
in Pb+Pb reactions at CERN SPS satisfy the coalescence model predictions, 
while they are known to be not satisfied by data from p + Pb
reactions at CERN SPS. Based on considerations of the
production ratios of strange hadrons,
 we have obtained an elementary,
self-consistent and model independent proof,
that {\it quark degrees of freedom are 
liberated} {\it in Pb + Pb reactions at CERN SPS} 
before the onset of hadron formation. 

\section*{Acknowledgement}
This work was supported by the OTKA Grant No. T029158.
 


\begin{thebibliography}{RS}
\bibitem{Bial}  
A. Bialas, Phys. Lett. {\bf B442} (1998) 449.
\bibitem{bz} T. S. Bir\'o and J. Zim\'anyi,
Nucl. Phys. {\bf A395} (1983) 525.
\bibitem{ALCOR} T.S. Bir\'o, P. L\'evai, and  J. Zim\'anyi,
Phys. Lett. {\bf B347} (1995) 6.
\bibitem{ALC97} T.S. Bir\'o, P. L\'evai, and J. Zim\'anyi,
J. Phys. {\bf G23} (1997) 1941.
\bibitem{ALC99} T.S. Bir\'o, P. L\'evai, and J. Zim\'anyi,
J. Phys. {\bf G25} (1999) 311. 
\bibitem{ALC96} J. Zim\'anyi, T.S. Bir\'o, T. Cs\"org\H o, and P. L\'evai,
Heavy Ion Phys. {\bf 4} (1996) 15.
\bibitem{EXP1} M. Kaneta et al., NA44 collaboration,
J. Phys. {\bf G23} (1997) 1865.
\bibitem{EXP2} R. Caliandro et al., WA97 collaboration,
J. Phys. {\bf G25} (1999) 171.
\bibitem{EXP3} C. Bormann for the NA49 Coll.,
J. Phys. {\bf G23} (1997) 1817.
\bibitem{scl99} A. Ster, T. Cs\"org\H o and B. L\"orstad,
hep-ph/9907338, Nucl. Phys. {\bf A} (1999) in press.
\bibitem{ALCOR2} T.S. Bir\'o, P. L\'evai, J. Zim\'anyi,
Phys. Rev. {\bf C59} (1999) 1574.
 
\end{thebibliography}
 \end{document}